\documentclass[twocolumn]{aastex701}

\newcommand{\mgi}{Mg\,{\sc ii}}
\newcommand{\wmgi}{W_{\rm Mg\,II}}
\newcommand{\ew}{\wmgi^{\rm corr}}
\newcommand{\wzero}{W_0}
\newcommand{\rp}{r_{\rm p}}

\newcommand{\ang}{\mbox{\AA}}

\shorttitle{Azimuthal \mgi\ Absorption around DESI LRGs}
\shortauthors{Wu \& Li}

\submitjournal{ApJ}

\begin{document}

\title{Evidence for Azimuthally Anisotropic \mgi\ Absorption around DESI Luminous Red Galaxies}

\author{Xuanyi Wu}
\affiliation{Department of Astronomy, Tsinghua University, Beijing 100084, China}
\affiliation{China Unicom Data Intelligence Co., Ltd., Hangzhou 311215, China}
\email{}

\author[0000-0002-8711-8970]{Cheng Li}
\affiliation{Department of Astronomy, Tsinghua University, Beijing 100084, China}
\email{cli2015@tsinghua.edu.cn}

\correspondingauthor{Cheng Li}
\email{cli2015@tsinghua.edu.cn}

\begin{abstract}
We use luminous red galaxies (LRGs) and background quasar spectra from DESI DR1 to measure the mean \mgi\ equivalent width around massive quiescent galaxies as a function of projected radius and azimuth relative to the projected major axis. Our forced-measurement approach assigns a \mgi\ doublet-window EW to every LRG--quasar pair, including spectra without individually detected absorbers, and subtracts a redshift-matched random-control signal measured from the same normalized quasar spectra. The all-angle profile declines smoothly over $\rp=0.01$--$1.0$ proper Mpc and shows stronger inner absorption at $0.4<z_{\rm LRG}<0.75$ than at $0.75<z_{\rm LRG}<1.1$. Superposed on this radial and redshift dependence, sightlines near the LRG major axis show enhanced absorption relative to the minor axis at $\rp\simeq30$--$80$ kpc. For the full $0.4<z_{\rm LRG}<1.1$ sample, the integrated fiducial major-minus-minor difference over $\rp=0.03$--$0.077$ Mpc is $0.184\pm0.075\,\ang$, with a position-angle randomization probability $p=0.013$. The redshift split shows no significant evolution in the anisotropy amplitude, even though the inner EW profile itself evolves. Our results provide evidence for a localized major-axis enhancement of \mgi-bearing cool/warm gas around LRGs, showing that azimuthal CGM structure is measurable in massive quiescent halos as well as in star-forming systems.
\end{abstract}


\section{Introduction} \label{sec:intro}

Quasar absorption lines have long provided a way to detect diffuse gas around galaxies that is otherwise difficult to observe directly \citep{BahcallSpitzer1969,Bergeron1986,Steidel1994}. In the modern circumgalactic-medium (CGM) picture, such gas records the exchange of baryons and metals between galaxies and their environments \citep[e.g.,][]{Tumlinson2017,Peroux2020,Faucher-Giguere2023}. The \mgi\ doublet is especially useful at intermediate redshift because it is strong, accessible to optical surveys, and sensitive to chemically enriched gas at $T\sim10^4$ K. A persistent result from both targeted and statistical studies is that massive, mostly passive galaxies retain non-negligible \mgi-bearing gas in their halos \citep[e.g.,][]{BowenChelouche2011,Huang2016,Zhu2014,Chen2018,Zahedy2019,Lan2020}. Understanding the geometry of this gas is a route toward identifying its origin.

Azimuthal-angle measurements are particularly diagnostic. In star-forming disk galaxies, enhanced absorption along minor axes is commonly interpreted as evidence for bipolar outflows, while major-axis enhancement can point toward extended disks, accretion, satellites, or filamentary structure \citep[e.g.,][]{Bordoloi2011,Kacprzak2012,Nielsen2016,Lan2018}. Emission-line galaxy samples observed with DESI show related minor-axis signatures in \mgi\ absorption \citep{Chen2025}. For luminous red galaxies (LRGs), however, the interpretation is less settled: their low star-formation rates weaken the classic outflow picture, while their massive halos and satellite populations provide several plausible channels for anisotropic cool gas.

\begin{table*}
    \centering
    \caption{Accepted LRG--QSO pair counts by radial bin and sample selection. Each entry lists fiducial / $S/N>0$ / $S/N\ge3$ accepted pair measurements after the common LRG-pair/random-control measurement, wavelength-mask, and EW-quality requirements. The fiducial sample further applies the continuum-quality cut described in Section~\ref{sec:data}.}
    \label{tab:sample}
    \scriptsize
    \begin{tabular}{lccc}
    \hline
    $\rp$ [Mpc] & $0.4<z_{\rm LRG}<1.1$ & $0.4<z_{\rm LRG}<0.75$ & $0.75<z_{\rm LRG}<1.1$ \\
    \hline
    0.010--0.030 & 121 / 197 / 92 & 86 / 134 / 66 & 35 / 63 / 26 \\
    0.030--0.050 & 308 / 558 / 239 & 212 / 391 / 173 & 96 / 167 / 66 \\
    0.050--0.077 & 711 / 1,262 / 549 & 443 / 820 / 356 & 268 / 442 / 193 \\
    0.077--0.130 & 2,336 / 4,055 / 1,761 & 1,504 / 2,686 / 1,170 & 832 / 1,369 / 591 \\
    0.130--0.215 & 6,833 / 11,774 / 5,084 & 4,446 / 7,812 / 3,435 & 2,387 / 3,962 / 1,649 \\
    0.215--0.360 & 23,471 / 40,658 / 17,564 & 15,286 / 27,183 / 11,812 & 8,185 / 13,475 / 5,752 \\
    0.360--0.550 & 57,956 / 100,335 / 42,921 & 36,797 / 65,657 / 28,307 & 21,159 / 34,678 / 14,614 \\
    0.550--0.770 & 108,822 / 189,484 / 80,596 & 69,127 / 124,202 / 53,063 & 39,695 / 65,282 / 27,533 \\
    0.770--1.000 & 167,962 / 291,708 / 124,072 & 105,625 / 188,791 / 81,116 & 62,337 / 102,917 / 42,956 \\
    Total & 368,520 / 640,031 / 272,878 & 233,526 / 417,676 / 179,498 & 134,994 / 222,355 / 93,380 \\
    \hline
    \end{tabular}
\end{table*}

LRGs are also a natural population in which to search for anisotropic halo gas because their stellar major axes are correlated with the surrounding matter distribution. SDSS studies show that satellite galaxies are preferentially distributed along the major axes of central galaxies, with strong signals for red systems, and that LRG shapes are intrinsically aligned with the large-scale density field \citep{Yang2006,Hirata2007,Okumura2009,Lamman2023}. Related SDSS and Millennium Simulation analyses show that red, luminous galaxies preferentially align with the large-scale distribution of neighboring galaxies, with the signal especially strong for LRG-like systems \citep{Faltenbacher2009}. If \mgi-bearing gas around LRGs is supplied or redistributed by satellites, tidal debris, or filamentary environments, the projected galaxy orientation may therefore be linked to the absorber distribution.

DESI now provides a sample large enough to move beyond detected-absorber counts and measure the mean \mgi\ equivalent-width (EW) field around LRGs directly. Earlier SDSS cross-correlation and close-pair studies established that \mgi\ absorbers and LRG environments are connected from inner-halo to group scales \citep{Bouche2006,Lundgren2009,Gauthier2009,Gauthier2010,BowenChelouche2011}. Subsequent SDSS work found extended cool gas around LRGs and either weak or population-dependent azimuthal trends in absorber covering fraction \citep{Zhu2014,Huang2016,Anand2021,Lan2020}. DESI studies have since measured the radial, kinematic, and host-property dependence of \mgi\ absorption around LRGs \citep{Chang2024,Chang2025Y1}, while \citet{Chen2025} did not find a clear LRG azimuthal signal in their detected-absorber analysis. This paper presents DESI DR1 forced-EW evidence for a localized major-axis enhancement in the mean \mgi\ EW field of LRG halos, tested directly with position-angle randomizations. We measure the profile at $\rp<1$ Mpc over $0.4<z<1.1$, examine the redshift dependence of both the total EW profile and the anisotropy, and use continuum-quality variations as consistency tests.

Throughout this paper, projected distances are proper distances computed with the Planck18 cosmology \citep{Planck2020}, with $H_0=67.66\,{\rm km\,s^{-1}\,Mpc^{-1}}$, $\Omega_{\rm m}=0.30966$, $\Omega_{\Lambda}=0.68885$, and $\Omega_{\rm b}=0.04897$.

\section{Data and Measurement} \label{sec:data}

\subsection{DESI LRG--QSO Pairs}

We use DESI DR1 LRG and quasar catalogs \citep{DESI2016a,DESI_instr,DESI_DR1}. Foreground LRGs are restricted to the redshift range of $0.4<z_{\rm LRG}<1.1$ and are required to have usable projected position angles from Legacy Survey imaging \citep{Dey2019}. The position angle is computed from the Legacy Surveys/Tractor galaxy-model ellipticity components, $\mathrm{SHAPE\_E1}$ and $\mathrm{SHAPE\_E2}$, and is defined modulo $180^\circ$; this model shape is not a separate single-band position-angle measurement. The corresponding minor-to-major axis ratio is used only for the robustness test in Section~\ref{sec:robust}. Quasars are required to lie behind the LRGs, with a redshift buffer of $\Delta z>0.1$, and to have finite DESI redshifts. We construct LRG--QSO pairs with projected separation $0.01<\rp<1.0$ Mpc, where the lower bound avoids the regime in which the foreground galaxy light and fiber-positioning geometry can dominate the sightline selection. Pairs are then filtered by the spectral-quality and random-control requirements described below.

Table~\ref{tab:sample} summarizes the accepted pair counts supporting the fiducial and continuum-quality comparison measurements. The number of accepted pairs rises steeply with radius because the analysis is based on projected LRG--QSO geometry rather than on pre-detected absorbers. The innermost bins therefore carry the largest statistical uncertainties, but they also provide the cleanest test of whether any azimuthal dependence is localized close to the LRG.

\subsection{Forced Mg II Equivalent Widths}\label{sec:measurement_method}

We measure \mgi\ absorption with a forced-EW method rather than by first identifying individual absorbers. The goal is to measure the mean absorber field associated with a foreground population, not the incidence of features that pass a line-detection threshold. Every geometrically selected LRG--QSO pair therefore receives an EW estimate at the LRG redshift. Positive, negative, and low-significance measurements are all retained until the common quality cuts are applied. In this respect the estimator is analogous to a stacked measurement, but it is carried out at the pair level. This makes the same measurement naturally available in radial, redshift, orientation, and quality subsamples, and allows bootstrap, jackknife, and random-control operations to be applied to the same pair list. The same forced-EW and random-subtraction framework described here has also been applied to DESI \mgi\ measurements in \citet{Wu2024}.

For each background quasar, we use the DESI DR1 coadded spectrum on the merged wavelength grid, covering $3600$--$9824\,\ang$. We first model the broad quasar continuum in the quasar rest frame with non-negative matrix factorization (NMF; \citealt{Lee1999}). For each quasar-redshift interval we use fixed NMF bases and solve for non-negative coefficients with inverse-variance weights, masked pixels, and missing data following the heteroscedastic NMF formalism of \citet{Zhu2016}. After division by the NMF continuum, we apply an iterative median residual correction to remove broad residual structure. We then use local sidebands around the absorber-frame \mgi\ doublet to make a small local continuum adjustment and to quantify the continuum quality in the immediate wavelength region used for the EW measurement. This local step is important because the statistical signal is obtained by averaging many weak pair-level measurements; even small continuum residuals can otherwise bias the mean.

We shift the normalized spectrum to the LRG rest frame and measure the rest-frame EW in a fixed \mgi\ doublet window:
\begin{equation}\label{eqn:W0i}
    W_{0,i} =
    \sum_{\lambda_{\rm r}=\lambda_l}^{\lambda_u}
    \left[1-f_{{\rm norm},i}(\lambda_{\rm r})\right]\Delta\lambda_{\rm r}.
\end{equation}
The window limits are $\lambda_l=2792.6\,\ang$ and $\lambda_u=2807.3\,\ang$, enclosing both members of the \mgi\ doublet; throughout this paper we refer to the corresponding corrected, binned quantity as $\wmgi$. Observed-frame wavelength intervals affected by strong sky residuals are masked for both the LRG-pair and random-control measurements: $4300$--$4450\,\ang$, $5550$--$5605\,\ang$, and $5720$--$5865\,\ang$. A pair is rejected if the \mgi\ window overlaps these masked intervals or if too few valid pixels remain to define the window.

Equation~(\ref{eqn:W0i}) defines the forced EW for an individual LRG--QSO pair. For a radial, redshift, and azimuthal bin $B$, we then average the accepted pairs and form the corrected profile as
\begin{equation}\label{eqn:Wcorr}
    \ew(B)
    \equiv
    \left\langle \wzero\right\rangle_{\rm cor}(B)
    =
    \left\langle \wzero\right\rangle_B
    -
    \left\langle W_{0,{\rm ran}}\right\rangle_B .
\end{equation}
Both terms in equation~(\ref{eqn:Wcorr}) are weighted averages over the same accepted pairs in $B$, defined by
\begin{equation}
    \left\langle W_0\right\rangle_B
    =
    \frac{\sum_{i\in B}\omega_i W_{0,i}}{\sum_{i\in B}\omega_i},
\end{equation}
and
\begin{equation}
    \left\langle W_{0,{\rm ran}}\right\rangle_B
    =
    \frac{\sum_{i\in B}\omega_i \overline{W}_{0,{\rm ran},i}}{\sum_{i\in B}\omega_i}.
\end{equation}
For the pair catalog used in this work, $\omega_i=1$. The random-control quantity for pair $i$, $\overline{W}_{0,{\rm ran},i}$, is the mean EW from the valid random foreground redshifts assigned to that pair. These random redshifts are drawn to follow the LRG-redshift selection while being detached from the true LRG positions. They are required to keep the quasar behind the foreground redshift, to place the \mgi\ window redward of the quasar Ly$\alpha$ forest, and to satisfy the same observed-frame wavelength masks and local continuum-quality requirements as the LRG-pair measurement. We require at least three valid random placements for each accepted pair. Because the same weights and quality masks enter both averages, equation~(\ref{eqn:Wcorr}) is equivalent to averaging the pair-level differences $W_{0,i}-\overline{W}_{0,{\rm ran},i}$. The subtraction removes residual spectral, continuum, and redshift-dependent systematics that are not associated with the foreground LRGs.

\begin{figure}
    \centering
    \includegraphics[width=\columnwidth]{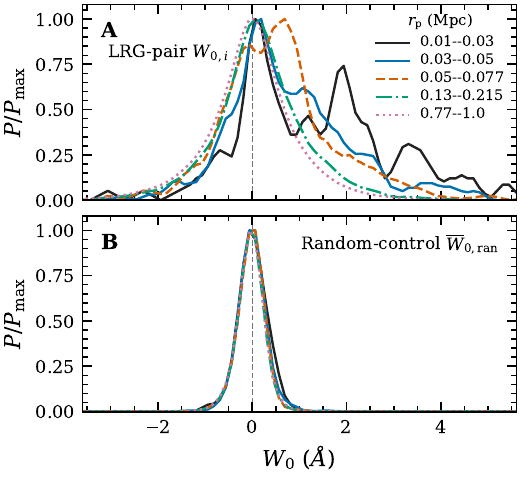}
    \caption{Pair-level forced-EW distributions for the fiducial $0.4<z<1.1$ sample. The upper panel shows the uncorrected LRG-pair measurements $W_{0,i}$ in selected radial bins. The lower panel shows the corresponding pair-level random-control means $\overline{W}_{0,{\rm ran},i}$. Each curve is normalized by its own maximum to emphasize the distribution shape.}
    \label{fig:hist}
\end{figure}

\begin{figure}
    \centering
    \includegraphics[width=\columnwidth]{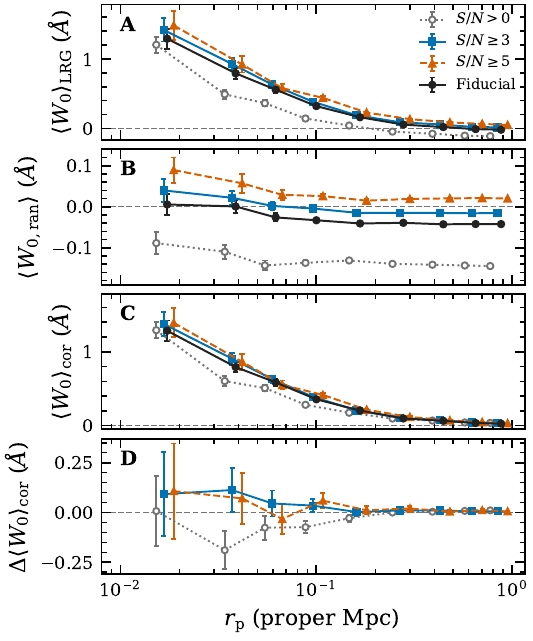}
    \caption{Continuum-quality comparison for the all-angle $0.4<z<1.1$ sample. From top to bottom, the panels show the uncorrected LRG-pair \mgi-window EW, the redshift-matched random-control EW, the corrected EW, and the corrected-EW difference relative to the fiducial profile. The EW is integrated over the rest-frame $2792.6$--$2807.3\,\ang$ doublet window. Points are shifted slightly in $\rp$ for readability.}
    \label{fig:method}
\end{figure}

Figure~\ref{fig:hist} illustrates the forced-measurement idea at the pair level. The LRG-pair $W_{0,i}$ distributions are broad because individual sightlines are noisy and absorption varies from pair to pair, but their positive tail and mean shift are strongest at small $\rp$ and weaken toward larger projected separations. The matched random-control distributions remain narrower and are centered close to zero, showing the scale-independent spectral residual pattern that is removed by equation~(\ref{eqn:Wcorr}). The measurement therefore does not require the individual-pair EW distribution to reveal a visually obvious absorber population; the signal is the difference between the mean LRG-pair distribution and the mean random-control distribution.

We adopt a fiducial continuum-quality selection rather than using either all positive-S/N spectra or only high-S/N spectra. Tests on DESI spectra showed that a permissive $S/N>0$ cut maximizes the number of pairs but can admit poorly normalized spectra with residual continuum or sky-subtraction structure, producing biased corrected EWs in some radial bins. The problematic cases are concentrated among low-S/N spectra with large local continuum scatter in the \mgi\ sidebands. A stricter $S/N\ge3$ cut is cleaner but discards many otherwise usable spectra and increases the statistical errors. Our fiducial sample therefore keeps pairs with either continuum $S/N\ge3$ or $0.5\le S/N<3$ when the local sidebands are smooth, quantified by local continuum scatter $\le0.5$. This choice is intended to preserve unbiased mean EWs while retaining much of the statistical power of the low-S/N data. As expected, the fiducial errors are smaller than those from the $S/N\ge3$ sample but larger than those from the less restrictive $S/N>0$ sample. We also require finite LRG-pair and random-control EW measurements, at least three valid random placements, and pair-level $|W_{0,i}-\overline{W}_{0,{\rm ran},i}|<5\,\ang$.

Figure~\ref{fig:method} illustrates the continuum-quality comparison for the full $0.4<z<1.1$ sample. As can be seen from the upper two panels, the uncorrected LRG-pair and random-control profiles vary with continuum-quality selection, as expected when low-S/N spectra are included. After the random subtraction, as shown in the third panel, the corrected profiles from the fiducial, $S/N\ge3$, and $S/N\ge5$ samples agree well within the measurement errors, while the profile from the $S/N>0$ sample is significantly lower at small projected separations ($\rp\lesssim0.2$Mpc). The differences can be more clearly seen from the bottom panel which shows the relative differences of the $S/N>0$, $S/N\ge3$ and $S/N\ge5$ samples with respect to the fiducial sample. In what follows, we obtain our primary measurements from the fiducial sample which balances continuum reliability and statistical precision, while the $S/N>0$ and stricter-S/N samples will be used for robustness checks.

\begin{figure*}
    \centering
    \includegraphics[width=0.9\textwidth]{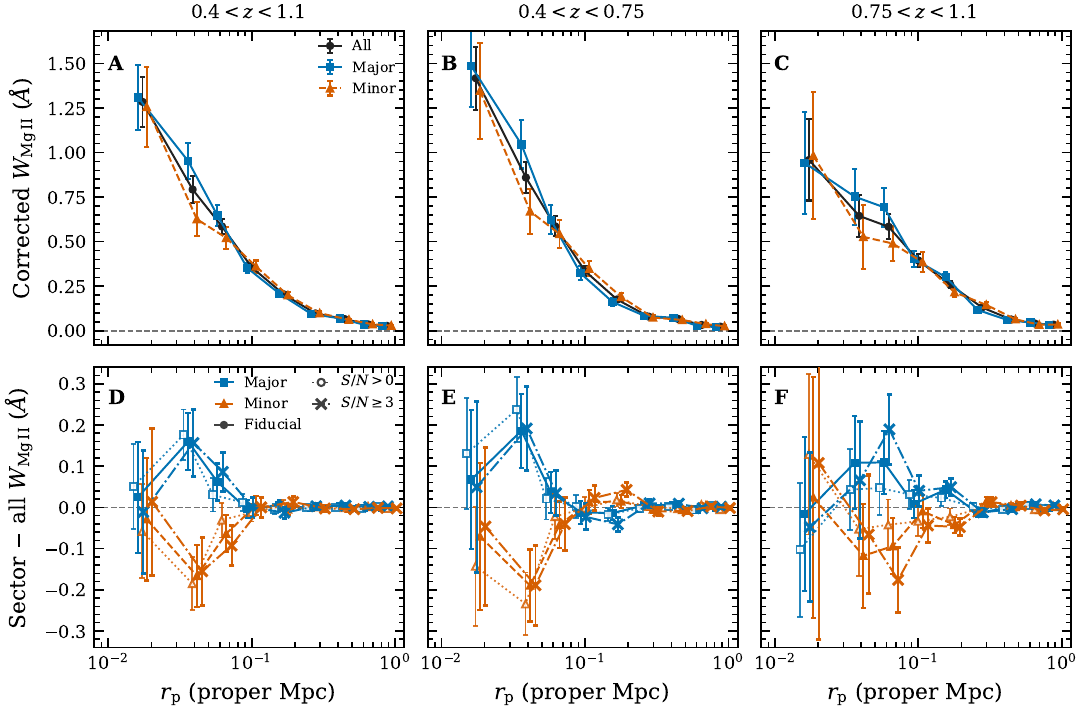}
    \caption{\mgi-window EW anisotropy around DESI LRGs at $\rp<1$ Mpc. Columns show the full $0.4<z<1.1$ sample and the two redshift bins split at $z=0.75$. The upper panels show the all-angle, major-axis, and minor-axis fiducial profiles. The lower panels show major-minus-all and minor-minus-all residuals; filled connected symbols are fiducial, open connected symbols show the $S/N>0$ comparison sample, and connected crosses show the $S/N\ge3$ comparison sample. Points are shifted slightly in $\rp$ for readability. Combining the full redshift range gives the tightest constraint, while the redshift split shows that the strongest single-bin contrast is in the low-redshift subsample.}
    \label{fig:combined}
\end{figure*}

\subsection{Azimuthal Sectors and Uncertainties}\label{sec:PA_method_test}

The azimuthal angle is measured between the projected LRG major axis and the LRG--QSO sightline, folded into $0^\circ\le|\Delta{\rm PA}|\le90^\circ$. We define the major-axis sector as $|\Delta{\rm PA}|\le45^\circ$ and the minor-axis sector as the complementary set. This division keeps the two orientation sectors similar in area and, after the pair-quality cuts, similar in pair count. We measure the all-angle, major-axis, and minor-axis EW profiles in logarithmic radial bins spanning $\rp=0.01$--$1.0$ Mpc.

Errors on individual radial bins are estimated with both spatial jackknife resampling and LRG-level bootstrap resampling. For the jackknife, the survey footprint is divided into 30 sky regions and each region is omitted in turn. This test is sensitive to field-to-field variations and to the possibility that a small number of sky regions dominate the signal. For the bootstrap, we resample the LRGs with replacement and recompute the pair-weighted profiles. This keeps all QSO sightlines associated with a resampled LRG together and therefore treats the foreground galaxies as the statistical units. Unless stated otherwise, the quoted errors on integrated amplitudes are bootstrap errors, while the jackknife is used as an independent check on the radial profiles and composite-stack comparison.

The main anisotropy statistic is the weighted mean of the major-minus-minor profile over a specified radial interval. We focus on $\rp=0.03$--$0.077$ Mpc because this is where the radial profiles show the strongest localized contrast and because it corresponds to the tens-of-kpc regime in which gas associated with satellites, stripping, or the inner halo is most plausibly tied to the galaxy orientation. To test whether the signal depends on the measured LRG position angles, we randomize the LRG position angles 3000 times and repeat the full sector assignment and integrated-amplitude measurement. We quote the resulting two-sided probability as the local PA-randomization probability for the chosen radial interval.

We also perform a look-elsewhere test over contiguous radial windows. For each PA-randomized catalog, we record the largest absolute signal-to-noise ratio over a predefined set of windows and compare this distribution with the observed maximum. We use three window families: all contiguous windows within $\rp=0.01$--$1.0$ Mpc, windows restricted to $\rp<0.215$ Mpc, and windows within $\rp=0.03$--$0.13$ Mpc. This test is not used to define the signal; it quantifies how unusual the largest apparent radial contrast would be if the LRG position angles were unrelated to the \mgi\ EW field.

\section{Results} \label{sec:results}

\subsection{A Localized Major-Axis Enhancement}

The upper-left panel of Figure~\ref{fig:combined} shows the all-angle, major-axis and minor-axis $\wmgi$ profiles measured from the fiducial sample. Overall, all the profiles decline rapidly with projected radius. The all-angle profile decreases from $1.28\,\ang$ at $\rp=0.01$--$0.03$ Mpc to $0.027\,\ang$ at $\rp=0.77$--$1.0$ Mpc. This radial decline is expected for \mgi-bearing gas in the inner circumgalactic medium. 

The profiles at the major and minor axes are similar to the all-angle profile over most of the radial bins. Significant azimuthal dependence is detected only at a few $\times10$ kpc, where the major-/minor-axis profile is above/below the all-angle profile. This difference is more clearly seen in the lower panel which shows the difference of the major-/minor-axis profiles relative to the all-angle profile. At the radial bin of $\rp=0.03$--$0.05$ Mpc, the $\wmgi$ along the major axis is enhanced by $0.160\pm0.070\,\ang$, while the minor-axis $\wmgi$ is reduced by $0.166\pm0.076\,\ang$. Equivalently, the direct major-minus-minor difference in this bin is $0.326\pm0.138\,\ang$. The signal remains ($W_{\rm major}-W_{\rm minor}=0.184\pm0.075\,\ang$) when integrating over the adjacent bins spanning $\rp=0.03$--$0.077$ Mpc. This corresponds to a bootstrap signal-to-noise ratio of 2.46. The position-angle randomization test gives a local two-sided null probability $p=0.013$, equivalent to 2.47 Gaussian sigma for this radial interval. 

The anisotropy is not a broad average over a wide radial range. A maximum-window PA test gives $p_{\rm global}=0.112$ within $\rp<0.215$ Mpc and $0.197$ across $\rp=0.01$--$1.0$ Mpc. When integrated over $\rp=0.01$--$1.0$ Mpc, the fiducial major-minus-minor amplitude is $-0.0007\pm0.0032\,\ang$. The evidence therefore points to a localized tens-of-kpc anisotropy rather than a scale-independent orientation dependence.

\subsection{Redshift Dependence of the EW Profile and Anisotropy}

We split the sample at $z=0.75$ to compare the low-redshift bin, $0.4<z<0.75$, with the higher-redshift bin, $0.75<z<1.1$. The middle and right columns of Figure~\ref{fig:combined} show the corresponding profiles and axis residuals. The all-angle profile is stronger at lower redshift in the innermost halo: the fiducial $\wmgi$ is $1.42\pm0.18\,\ang$ at $\rp=0.01$--$0.03$ Mpc and $0.86\pm0.09\,\ang$ at $\rp=0.03$--$0.05$ Mpc for $0.4<z<0.75$, compared with $0.96\pm0.23\,\ang$ and $0.64\pm0.12\,\ang$ in the same two radial bins for $0.75<z<1.1$. Beyond these innermost bins, the two profiles are closer and do not show a single monotonic ordering. Thus the clearest redshift evolution in this measurement is a stronger inner all-angle EW profile at lower redshift.

The anisotropy amplitude shows weaker redshift dependence than the EW profile itself. The low-redshift bin shows the clearest single-bin major-axis enhancement at $\rp=0.03$--$0.05$ Mpc, with $W_{\rm major}-W_{\rm minor}=0.376\pm0.174\,\ang$ in the fiducial sample. Due to larger uncertainties in the innermost bins, the higher-redshift bin shows an anisotropy signal that is similar to but not as significantly as in the lower-redshift bin. Within these uncertainties, the redshift-split measurements are consistent with a common anisotropy amplitude.

\section{Discussion} \label{sec:discussion}

\subsection{Comparison with Previous Work}

Our measurement adds two pieces of information to previous studies of cool gas around massive galaxies. First, the forced-EW profile shows stronger inner \mgi\ absorption at $0.4<z<0.75$ than at $0.75<z<1.1$. The existence of cool gas around massive quiescent galaxies is well established by SDSS clustering, close-pair, stacking, and absorber-catalog measurements \citep{Bouche2006,Lundgren2009,Gauthier2009,Gauthier2010,BowenChelouche2011,Zhu2014,Huang2016,Anand2021,Lan2020}, and by recent DESI analyses \citep{Chang2024,Chang2025Y1}. Our measurement is complementary because it averages the \mgi\ doublet-window EW over all selected LRG--QSO pairs, including nondetections, and subtracts a matched random-control signal. It therefore probes the mean inner-halo absorption field rather than only the incidence of individually detected absorbers.

Second, we find evidence for a localized azimuthal dependence in the LRG \mgi\ EW field. This differs from the better-established minor-axis \mgi\ enhancement around star-forming or emission-line galaxies, usually interpreted as outflow-related gas \citep{Bordoloi2011,Kacprzak2012,Nielsen2016,Lan2018,Chen2025}. It also differs from previous LRG work, where azimuthal signals were weak, scale-limited, or not significantly detected \citep{Huang2016,Chen2025}. The sign of the DESI LRG signal is especially informative: the enhancement is along the projected major axis, not the minor axis. This points to a geometry more closely connected to the stellar body, satellite distribution, or surrounding environment of massive quiescent galaxies than to the bipolar outflow geometry often discussed for star-forming disks.

\begin{figure*}
    \centering
    \includegraphics[width=0.9\textwidth]{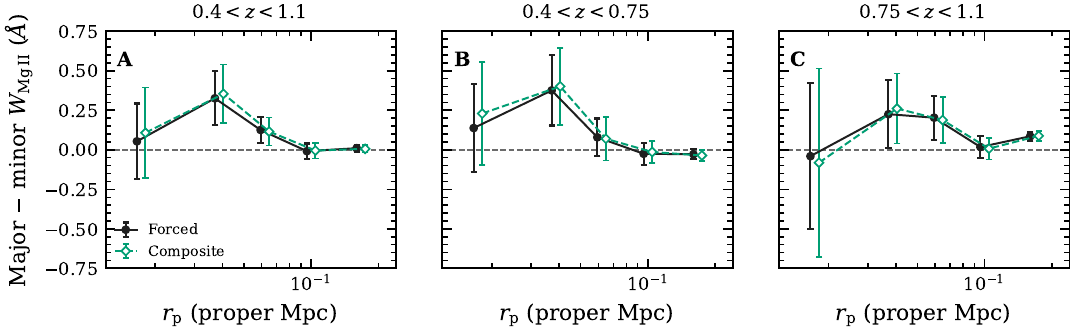}
    \caption{Composite-stack comparison for the combined and redshift-split $\rp<1$ Mpc measurements. The panels compare the pair-level forced estimator with a conventional composite-spectrum measurement in the same radial, orientation, and redshift bins. Both measurements are corrected by matched random-control subtraction, and the error bars are spatial jackknife estimates.}
    \label{fig:composite}
\end{figure*}

\subsection{Implications for Cool Gas in Massive Halos}

The major-axis \mgi\ enhancement is qualitatively consistent with the known alignment between luminous red galaxies and the surrounding galaxy distribution. SDSS group studies show that satellite galaxies are preferentially distributed along the projected major axes of central galaxies, especially for red central and satellite populations \citep{Yang2006}, while intrinsic-alignment measurements show that LRG orientations trace the large-scale density field \citep{Hirata2007,Okumura2009,Lamman2023}. \citet{Faltenbacher2009} further showed that red, luminous SDSS galaxies have an excess of neighboring galaxies along their projected major axes, and that comparable alignments arise in simulations when central-galaxy orientations follow the inner halo. If satellites and large-scale structure are preferentially distributed along the LRG major axis, then cool gas associated with satellites, stripping, tidal debris, or anisotropic accretion could naturally produce a major-axis excess in the inner halo.

The radial localization is also important. The all-angle \mgi\ profile is smooth and declines strongly with radius, while the major--minor contrast averages to zero over the full $\rp<1$ Mpc range. This combination argues against a simple sector-wide normalization difference and instead favors a localized redistribution of cool gas around the mean profile. Such gas could be supplied by satellite material, stripped interstellar gas, residual cosmological inflow, or condensation from the hot halo, all of which have been discussed in theoretical models of cool gas in massive halos \citep[e.g.,][]{Keres2005,DekelBirnboim2006,Afruni2019,Voit2021}. These measurements do not distinguish among those channels. The inner all-angle EW profile evolves more clearly with redshift than the anisotropy amplitude, suggesting that the total amount or covering of cool/warm gas in the inner LRG halo and the angular redistribution of that gas may not evolve in lockstep.

\begin{figure*}
    \centering
    \includegraphics[width=0.9\textwidth]{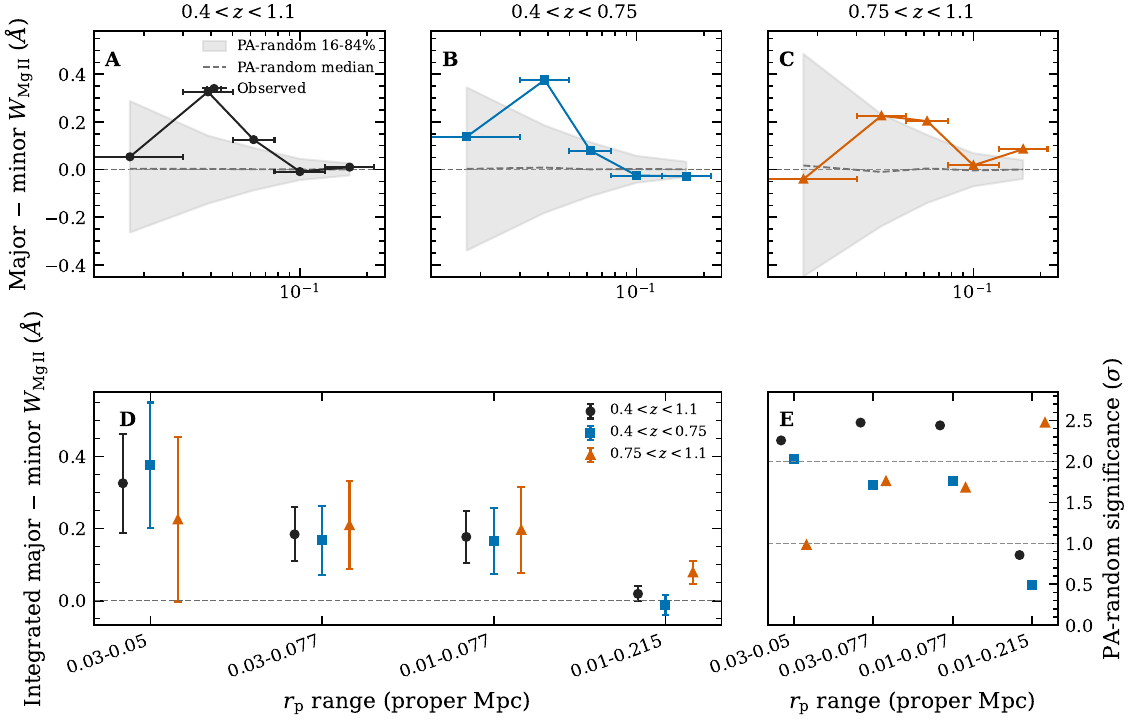}
    \caption{Position-angle randomization tests for the combined and redshift-split $\rp<1$ Mpc measurements. The upper panels compare the observed major-minus-minor profiles with measurements from randomized LRG position angles. The lower panels summarize the integrated major-minus-minor amplitude and the corresponding local two-sided PA-randomization significance for the full sample and for the two redshift bins.}
    \label{fig:panull}
\end{figure*}

\subsection{Robustness of the EW measurements} \label{sec:robust}

Measurements of $\wmgi$ profiles may be affected by wavelength-dependent spectral residuals and continuum-quality selection. In our work, these possible systematics are addressed by construction. We apply the same observed-frame sky masks to the LRG-pair and random-control measurements, so masked wavelength intervals cannot introduce a mismatch between the two terms in equation~(\ref{eqn:Wcorr}). The redshift-matched random subtraction then removes residual spectral features that are tied to the observed quasar spectra or to the redshift distribution rather than to the foreground LRGs.

Our forced measurements are robust to continuum-quality selection. As shown in Figure~\ref{fig:method}, although the uncorrected LRG-pair and random-control profiles change with the S/N selection, after random subtraction the fiducial, $S/N>0$, $S/N\ge3$, and $S/N\ge5$ profiles agree over most of the radial range. The $S/N>0$ comparison sample also shows a qualitatively similar inner-bin anisotropy, with a stronger single-bin signal at $\rp=0.03$--$0.05$ Mpc and a consistent integrated signal at $\rp=0.03$--$0.077$ Mpc. We do not use the $S/N>0$ sample as the fiducial measurement because tests of the low-S/N tail show that spectra with large local continuum scatter can bias the corrected EW. Conversely, the $S/N\ge3$ sample is conservative but noisier. The adopted fiducial selection is therefore a compromise between continuum reliability and statistical precision.

In addition, we form conventional composite spectra from the same locally normalized LRG-pair sightlines in each radial, orientation, and redshift bin, and subtract the matched random-control mean used for the forced estimator. If the weights and masks are matched, the composite and forced estimators should recover the same mean EW; their difference is mainly procedural rather than physical. The result is shown in Figure~\ref{fig:composite}. We see that the composite major-minus-minor profiles well reproduce the forced-EW profiles within the spatial-jackknife uncertainties. 

\begin{table}
    \centering
    \caption{Full-sample PA-randomization window-trial results. The search is over contiguous radial windows in the listed domain.}
    \label{tab:window_trials}
    \scriptsize
    \begin{tabular}{lccc}
    \hline
    Search domain & $N_{\rm win}$ & Best window [Mpc] & $p_{\rm global}$ \\
    \hline
    $0.03<\rp<0.13$ Mpc & 6 & 0.030--0.077 & 0.067 \\
    $\rp<0.215$ Mpc & 15 & 0.030--0.077 & 0.112 \\
    $0.01<\rp<1.0$ Mpc & 45 & 0.030--0.077 & 0.197 \\
    \hline
    \end{tabular}
\end{table}

For orientation-dependent $\wmgi$ measurements, possible systematics could be introduced by the uncertainty in the LRG position angle. To test this effect, as described in Section~\ref{sec:PA_method_test}, we perform a positional angle randomization test in which we randomize the LRG position angles 3000 times and repeat the measurements. The upper panels of Figure~\ref{fig:panull} compare the observed major-minus-minor profiles with the distribution obtained after randomizing the LRG position angles. As can be seen, the observed major-minus-minor axis $\wmgi$ is significantly stronger than the PA-random measurements in the radial bin of $0.03<\rp<0.05$ for both the full sample and the low-redshift subsample. The lower panels summarize the corresponding integrated amplitudes and local PA-randomization probabilities for the full sample and the two redshift bins. For the full sample, the local probability for the $\rp=0.03$--$0.077$ Mpc interval is $p=0.013$. Table~\ref{tab:window_trials} gives the corresponding look-elsewhere results for contiguous radial-window searches. The signal remains the highest-ranked full-sample window in each search family, but the global probabilities are larger than the local probability. We therefore regard $p=0.013$ as the local evidence for the physically motivated tens-of-kpc interval, and the global probabilities as a conservative measure of how strongly the data prefer this radial localization after allowing the window to vary.

\begin{table}
    \centering
    \caption{Axis-ratio robustness test for the full $0.4<z<1.1$ sample. The amplitude is $W_{\rm major}-W_{\rm minor}$ integrated over $\rp=0.01$--$0.077$ Mpc.}
    \label{tab:axis_ratio}
    \scriptsize
    \begin{tabular}{lccc}
    \hline
    LRG shape cut & $N_{\rm major}$ & $N_{\rm minor}$ & $\Delta W$ [$\ang$] \\
    \hline
    All & 584 & 556 & $0.171\pm0.073$ \\
    $b/a\le0.8$ & 510 & 474 & $0.145\pm0.079$ \\
    $b/a\le0.7$ & 408 & 364 & $0.125\pm0.091$ \\
    $b/a\le0.6$ & 283 & 254 & $0.148\pm0.109$ \\
    $b/a\le0.5$ & 179 & 145 & $0.150\pm0.144$ \\
    \hline
    \end{tabular}
\end{table}

Finally, we test whether rounder LRGs dilute the signal because their position angles are less precisely measured. Table~\ref{tab:axis_ratio} repeats the integrated inner-halo anisotropy measurement after imposing cumulative cuts on the LRG minor-to-major axis ratio ($b/a$). The pair counts decrease rapidly with stricter cuts, especially in the innermost radial bins. Within the enlarged uncertainties, the $b/a\le0.8$ and $b/a\le0.6$ samples remain consistent with the full fiducial sample. The stricter cuts do not reveal a hidden, much stronger anisotropy in highly elongated LRGs. Therefore, the measurement is not simply an average of a strong signal in elongated systems and noise from round systems. 

\section{Summary} \label{sec:summary}

In this work, we have measured forced-\mgi\ doublet-window EWs $\wmgi$ around luminous red galaxies (LRGs) from DESI DR1 as a function of azimuthal angle within $\rp<1$ Mpc. Our measurements are performed for the full sample over $0.4<z<1.1$, as well as two redshift bins split at $z=0.75$. We have performed several robustness tests to verify that our results are robust to the forced-measurement methodology and the uncertainties in LRG orientation. 

Our main conclusions can be summarized as follows.

\begin{enumerate}
    \item For $0.4<z<1.1$, the all-angle \mgi\ EW profile declines from $\sim1.3\,\ang$ at $\rp=0.01$--$0.03$ Mpc to $\sim0.03\,\ang$ at $\rp=0.77$--$1.0$ Mpc. Splitting at $z=0.75$ shows stronger inner EW at lower redshift.
    \item Over the same redshift range, the \mgi\ EW profile shows evidence for enhanced absorption along the projected LRG major axis at $\rp=0.03$--$0.077$ Mpc, with $W_{\rm major}-W_{\rm minor}=0.184\pm0.075\,\ang$. The anisotropy amplitude shows no significant redshift evolution.
    \item A position-angle randomization test gives a local $p=0.013$ for the integrated $\rp=0.03$--$0.077$ Mpc anisotropy, supporting an interpretation tied to the true LRG orientations. Window trials give $p_{\rm global}=0.112$--$0.197$.
    \item The anisotropy is localized to tens-of-kpc scales and is not detected after averaging over the full $\rp=0.01$--$1.0$ Mpc range.
\end{enumerate}

These results indicate an anisotropic distribution of \mgi-bearing cool/warm gas in the inner circumgalactic medium of massive DESI LRGs, superposed on an evolving mean EW profile. Together with previous DESI measurements of the radial and kinematic properties of LRG halo gas, they add the azimuthal dimension needed to connect the absorber distribution with the geometry of massive galaxies, their satellite environments, and the surrounding cosmic web. Distinguishing among possible gas origins will require satellite catalogs, LRG properties, and absorber kinematics.

\begin{acknowledgments}

This work is supported by the National Key R\&D Program of China (grant No. 2018YFA0404502), and the National Natural Science Foundation of China (grant No. 12433003). We thank the DESI Collaboration for making the DR1 spectroscopic data products and catalogs publicly available.
\end{acknowledgments}

\section*{Data Availability}

The DESI DR1 spectroscopic data products and Legacy Survey imaging catalogs used in this work are publicly available. The derived measurements underlying this article will be shared on reasonable request to the corresponding author.

\software{Astropy, NumPy, SciPy, Matplotlib, h5py}

\bibliographystyle{aasjournalv7}
\bibliography{references}

\end{document}